\documentstyle[rmaaconf]{article}
\input tex.def
\input psfig.tex

\begin{document}

%
%

\title{YOUNG STAR CLUSTERS IN STARBURST ENVIRONMENTS\altaffilmark{1}}

\author{Luis C. Ho\altaffilmark{2}}
\altaffiltext{1}{Invited paper to appear in {\it Rev. Mex. Astr. Astrofis.} 
(1996), proceedings of {\it Starburst Activity in Galaxies}, ed. J. Franco, 
R.~Terlevich, \& G.~Tenorio-Tagle.}
\altaffiltext{2}{Harvard-Smithsonian Center for Astrophysics, 60 Garden St., MS-42, Cambridge, MA 02138, U.S.A.}

\begin{resumen}
        Unas recientes observaciones de alta resoluci\'on llevadas a cabo por 
el {\it Telescopio Espacial Hubble (HST)} revelan la presencia frecuente de 
c\'umulos estelares muy compactos de luminosidad extrema (``superc\'umulos 
estelares'') en sistemas con brotes de formaci\'on estelar reciente. El modo
de formaci\'on en c\'umulos parece ser el dominante en los brotes de 
formaci\'on estelar. Se resumen las propiedades principales de los c\'umulos
j\'ovenes. Un nuevo estudio global de im\'agenes ultravioletas de las regiones
centrales de galaxias cercanas realizado por el {\it HST} indica que los 
c\'umulos j\'ovenes se forman en una variedad muy amplia de entornos. En 
concreto, los anillos circumnucleares de formaci\'on estelar poseen un buen
n\'umero de c\'umulos y aqu\'{\i} se presentan varios ejemplos obtenidos a 
partir de im\'agenes recientes. Se ha especulado acerca de la posibilidad de 
que estos superc\'umulos estelares sean los equivalentes actuales de los
c\'umulos globulares j\'ovenes. Presentar\'e evidencias que sugieren que al
menos algunos superc\'umulos estelares tienen masas y densidades de masa 
comparables a las de los c\'umulos globulares actuales de la V\'{\i}a
L\'actea.
\end{resumen}

\begin{abstract}
Recent high-resolution observations with the {\it Hubble Space Telescope (HST)}
reveal that young star clusters of extraordinary luminosity and compactness 
(``super star clusters'') are commonly found in starburst systems.  Cluster 
formation appears to be a dominant mode of star formation in starbursts.  The 
principal properties of the young clusters are summarized.  A new ultraviolet 
{\it HST} imaging survey of the central regions of nearby galaxies indicates 
that young clusters form in a wide range of environments.  Circumnuclear 
star-forming rings, in particular, are richly populated with clusters, and 
several examples from recent imaging studies are discussed.  There has been 
much speculation that super star clusters represent present-day analogs 
of young globular clusters.  I will present evidence suggesting that at least 
some super star clusters indeed have masses and mass densities comparable to 
those of evolved globular clusters in the Milky Way.

\end{abstract}

\section{Introduction}
In the last few years, {\it Hubble Space Telescope (HST)} imaging studies of a 
variety of extragalactic star-forming systems have identified a widespread
new class of star clusters.  The compactness and high luminosities of these 
objects, coupled with their inferred youth, have stimulated speculation that 
they represent present-day analogs of young globular clusters.  Although the 
existence of a few such ``super star clusters'' (hereafter SSCs) had been 
known from previous ground-based studies (e.g., Arp \& Sandage 1985; Melnick, 
Moles, \& Terlevich 
1985; Lutz 1991), it took the resolving power of {\it HST} to demonstrate the 
prevalence of this phenomenon.  Such clusters appear to be found in a wide 
array of environments, ranging from nearby dwarf galaxies (O'Connell, 
Gallagher, \& Hunter 1994; Hunter, O'Connell, \& Gallagher 1994; O'Connell 
\etal 1995; Leitherer \etal 1996; Gorjian 1996), to more distant merging and 
interacting systems (Holtzman \etal 1992; Whitmore \etal 1993; Conti \& Vacca 
1994; Vacca 1994; Shaya \etal 1994; Whitmore \& Schweizer 1995; Meurer \etal 
1995), to circumnuclear star-forming rings (Benedict \etal 1993; Barth \etal 
1995, 1996; Bower \& Wilson 1995; Maoz \etal 1996a).  In fact, as I will discuss
later, it appears that the formation of compact clusters may be surprisingly 
commonplace in most regions of galaxies experiencing elevated levels of star 
formation, and need not be restricted to the most extreme starbursting 
environments.

This contribution will give an overview of the principal properties of SSCs, 
describe a new {\it HST} imaging survey and its applicability to the study 
of clusters, and summarize some recent work by my collaborators and myself 
pertaining to clusters in circumnuclear rings.  I will also present 
evidence based on ground-based observations that at least some of the SSCs 
have dynamical masses resembling those of evolved globular clusters seen 
in the Galaxy.

\section{Properties of the Clusters}

\subsection{Environments} 
The initial results from {\it HST} 
imaging studies give the impression that SSCs are only found in rather violent 
settings such as merging and interacting galaxies.  The first announcement, 
for example, was that of the well-known merging and/or cooling-flow galaxy 
NGC~1275 (Holtzman \etal 1992).  Two similar studies of merger systems quickly 
followed: numerous SSCs were found in the ``Antennae'' system 
(NGC~4038/NGC~4039; Whitmore \etal 1993) and in the ``Atoms for Peace'' galaxy 
(NGC~7252; Whitmore \& Schweizer 1995).  Likewise, Conti and collaborators 
(Conti \& Vacca 1994; Vacca 1994, 1996; Leitherer \etal 1996; Conti, 
Leitherer, \& Vacca 1996) are finding from their ultraviolet (UV) imaging 
program that SSCs invariably show up in Wolf-Rayet (W-R) galaxies, most of 
which are known to show signs of interaction.  The discovery of luminous, 
young clusters generated considerable excitement and rekindled interest in 
the idea that globular clusters might form in galaxy mergers (Burstein 1987; 
Schweizer 1987; Ashman \& Zepf 1992). 

While these studies certainly point to a 
plausible relation between galaxy interaction and the formation of SSCs, one 
must be careful in drawing any direct, causal connection between the two 
processes.  The cases cited above by no means represent an unbiased sample; in 
fact, they were, to a large extent, selected based on prior knowledge 
of their extreme characteristics.  It is well known that galaxy interactions 
can give rise to intense starburst activity (see contributions in Shlosman 
1994).  Similarly, W-R galaxies are among the earliest and most 
vigorous manifestations of starbursts (Conti 1991; Vacca \& Conti 1992).  
Thus, an equally viable interpretation is that cluster formation is an 
integral part of star formation in starbursts in general.  In as much as 
galaxy interactions can trigger starbursts, there is a high likelihood of 
finding SSCs in interacting systems; but such an association is indirect --- 
interactions are sufficient but not necessary for cluster formation.  That 
SSCs are also found in galaxies that are not obviously interacting 
(O'Connell \etal 1994; Meurer \etal 1995) supports this viewpoint, as does the 
tendency for some SSCs to be located in circumnuclear rings and other 
relatively quiescent environments, as will be described below. 

\subsection{Luminosities}
SSCs have earned their superlative title largely because of their 
very high luminosities, which in many cases surpass that of the R136 
cluster in the 30 Doradus complex in the Large Magellanic Cloud (LMC).  Some 
of the clusters, for instance, have absolute visual magnitudes exceeding --14 
to --15, whereas $M_V$ = --11.3 mag for R136 (O'Connell \etal 1994).  The
clusters will fade, of course, as they age, by an amount depending 
on their current age.  According to models of evolving stellar 
populations (e.g., Bruzual \& Charlot 1993), a 10-Myr old cluster 
will fade by 6 to 7 mag in $V$ after 10--15 Gyr.  

Meurer \etal (1995) define SSCs as having an absolute magnitude at 2200 \AA\ 
$\leq$ --14 (roughly corresponding to $M_V\,\leq$ --13 mag for a cluster 
with an age of 10 Myr); this criterion, however, is clearly arbitrary, for 
there exists a {\it continuum} of luminosities among clusters.  The 
luminosity function at the bright end can be approximated by a power law 
of the form $dN\,\propto\, L^{\alpha}dL$, where $\alpha\,\approx$ --2 
(Whitmore \etal 1993; Meurer \etal 1995; Meurer 1995; Vacca 
1996).  Although the luminosity function at the faint end is increasingly 
affected by detection incompleteness, it evidently preserves the same 
functional form and slope (Maoz \etal 1996a).  The luminosity function, 
however, as pointed out by van den Bergh (1995), does not resemble that of 
evolved globular clusters, which obey a nearly universal Gaussian function, 
but rather that of open clusters (e.g., van den Bergh \& Lafontaine 1984).  
Van den Bergh cites this as evidence that the young clusters are more closely 
related to open clusters instead of globular clusters.  Several factors, 
however, weaken van den Bergh's argument.  As noted by Larson (1993), a direct 
comparison between the luminosity function of young clusters and globular 
clusters is inappropriate since the two populations are viewed at very 
different stages: the former most likely have a heterogeneous mix of ages 
while the latter represent an evolved, much more uniform group.  Dynamical 
evolution undoubtedly will substantially modify the luminosity function of 
the young population, especially at the faint, low-mass end.  Meurer (1995) 
has shown that the luminosity function of the clusters in the Antennae 
system (Whitmore \etal 1993) can be modeled as a combination of a 
globular cluster mass function and continuous cluster formation.

\subsection{Ages}
The ages of SSCs span a wide range from as young as a few Myr to as much as 
several hundred Myr.  In the case of the UV knots seen in W-R galaxies, the 
presence of W-R stars constrains the ages to be 3--8 Myr (Meynet 1995).  
In some cases, the young ages have been independently confirmed by 
spectral-synthesis modeling of the UV spectrum (Conti \etal 1996; Leitherer 
\etal 1996).

The ages for the bulk of SSCs, however, have been derived from broad-band 
optical colors.  Ages from this method, unfortunately, are notoriously 
difficult to interpret, since the broad-band colors also depend on metallicity 
and reddening.
From the narrow range of colors and the apparent absence of a ``fading 
vector'' in the distribution of $V$ magnitude versus $V$--$R$ color, 
Holtzman \etal (1992) argued that the SSCs in NGC~1275 have a relatively 
small spread in ages, perhaps between 10 to 300 Myr.  It has subsequently 
been shown by Faber (1993), however, that the Holtzman \etal images were not 
sufficiently deep to allow this test to be performed unambiguously.  Faber's 
simulations, in fact, cannot rule out cluster ages as large as 1 to 5 Gyr. 
The age determinations for the clusters in NGC~7252 (30--500 Myr; 
Whitmore \etal 1993) and the Antennae (4--40 Myr; Whitmore \& Schweizer 
1995) suffer from similar ambiguity, since they are also based on 
photometric properties.

In principle, spectroscopic data can furnish more reliable age estimates, but, 
thus far, attempts to derive ages from optical spectra have not 
yielded superior results compared to those based on photometric methods.  
Schweizer \& Seitzer (1993) obtained spectra of the two brightest clusters in 
NGC~7252 and showed that the integrated light comes predominantly from 
late-A to mid-F stars.  But beyond this broad statement, no definitive 
conclusion could be drawn, since the spectral indices in the blue-red region 
were of limited use for dating the clusters.  The Mg{\it b} and Fe~I \lamb5270 
features, for example, give degenerate values of the age, while the Balmer 
absorption lines, in addition to being discrepant with model predictions for 
unknown reasons, can be fitted with a large range of ages spanning almost a 
factor of 100.  Zepf \etal (1995) performed essentially the same analysis 
for the the brightest cluster in NGC~1275.  

\subsection{Sizes}
Little is known about the detailed structure of the clusters.  Even with
the superior optics of the refurbished {\it HST}, the two SSCs in the
relatively nearby galaxy NGC~1569 ($d\,\approx$ 2.5 Mpc) still remain unresolved
(Leitherer, these proceedings).  Given these limitations, it is impossible to
characterize the size of the clusters by conventional parameters such as the
core or tidal
radius.  Instead, the observationally most straightforward parameter is the
half-light radius ($R_{\rm h}$), sometimes also referred to as the effective
radius.  $R_{\rm h}$ also has the advantage of being relatively insensitive to
evolutionary or environmental effects (van den Bergh, Morbey, \& Pazder 1991).
Unfortunately, even such a simple parameter is difficult to determine with
accuracy.  As discussed by Meurer \etal (1995), the measurement of cluster
radii from {\it HST} images is complicated by severe crowding and background
confusion in many cases, rendering the sizes highly uncertain for galaxies
more distant than a few Mpc.  Meurer \etal note a tendency to systematically
overestimate the sizes of distant clusters located on high-surface
brightness backgrounds.  In view of these caveats, the half-light radii of
well-resolved SSCs seem to fall comfortably within the range of Galactic
globular clusters, whose median $R_{\rm h}\,\approx$ 3 pc (van den Bergh \etal
1991).  The apparent tendency for the $R_{\rm h}$ distribution of SSCs
to be skewed toward somewhat larger sizes is probably not significant for the
reasons mentioned above (Meurer \etal 1995).  In particular, Meurer (1995)
has shown that the large radii reported by Whitmore \etal (1993) for the
clusters in the Antennae have probably been overestimated, thus obviating
van den Bergh's (1995) other major objection to the young globular cluster
hypothesis for the SSCs in this system.

The compact sizes of SSCs can be used as supporting evidence that the
clusters are most likely gravitationally bound.  Any conceivable form of 
perturbation that may lead to the dispersal of the stars will traverse the 
clusters on a timescale shorter than 1 Myr, whereas the estimated ages are 
in some cases up to two orders of magnitude larger.

\vfill\eject 
\subsection{Masses}
SSCs are thought to have masses at least as large as 10$^4$ \solmass, and 
more likely
between 10$^5$ to 10$^6$ \solmass.  Unfortunately, the estimated masses are 
highly uncertain because invariably they are derived from population synthesis 
models that depend on a large number of poorly constrained parameters.  To 
obtain the total cluster mass, one must generally adopt a slope for the 
stellar initial mass function (IMF) and extrapolate it to low, essentially 
unobserved, masses.  A more direct and reliable method to estimate the 
cluster masses will be described below.

\subsection{Metallicities}
The presence of compact clusters appears to be uncorrelated with the 
metallicity of the environment; this is an important factor that should be 
considered in models attempting to account for the formation of 
these objects.  A number of clusters are present in the dwarf galaxy 
I~Zw~18 (Meurer \etal 1995); with a gas-phase oxygen abundance of 1/50 that 
of the Sun (Dufour, Garnett, \& Shields 1988), it is the most metal-deficient 
galaxy known.  The brightest cluster in NGC~1275, on the other hand, evidently
has a metallicity roughly close to solar (Zepf \etal 1995), as might be 
expected considering that the host is a cD galaxy.  Finally, circumnuclear 
rings (\S\ 4), which occur almost exclusively in early-type spirals 
(e.g., Ho, Filippenko, \& Sargent 1996a), are well known to have high metal 
abundances, often exceeding solar (e.g., Storchi-Bergmann, Wilson, 
\& Baldwin 1996).


\section{The {\it HST} Ultraviolet Snapshot Survey}

While the {\it HST} imaging studies discussed up to now have opened a new 
vista on the importance of young star clusters, we still do not know how 
common the cluster phenomenon truly is.  Are SSCs only found in extreme 
environments such as mergers and very luminous starbursts, or do they also 
occur in more quiescent settings?  The examples given in \S\ 1 suggest a more 
widespread occurance, but it is difficult to be quantitative without 
access to a proper control sample. 

Fortunately, there are a number of snapshot imaging surveys of nearby galaxies 
being conducted with {\it HST}, and some of these may be useful for addressing 
the statistical properties of 
SSCs.  Here, I will briefly mention a recently completed UV imaging survey 
with {\it HST} (see Maoz \etal 1996b for details).  In brief, UV (F220W 
filter; effective wavelength $\sim$2270 \AA) images of the central 22\asec\ 
$\times$ 22\asec\ were obtained with the Faint Object Camera (FOC) for 110 
nearby galaxies.  The images were taken prior to the {\it HST} refurbishment 
mission, and hence suffer from the effects of spherical aberration.  The final 
pixel scale is 0\farcs0225, and the core of the point-spread function has a
full width at half maximum (FWHM) of $\sim$0\farcs05. The targets were selected 
randomly from a complete sample of 240 large ($D\,>$ 6$^{\prime}$) and nearby 
($cz\,<$ 2000 \kms) galaxies as listed in the UGC and ESO catalogs, and hence 
constitute a well-defined, unbiased group suitable for a variety of 
statistical studies.  The UV passband is especially useful for studying 
regions of recent star formation.

Even a cursory glance at the images in the Maoz \etal (1996b) catalog 
reveals that the UV emission exhibits a diverse assortment of morphologies.  
Of relevance in the present context are those images showing compact, 
pointlike sources.  
Figure 1 illustrates two examples.
NGC~3077 is a nearby amorphous galaxy interacting 
with M81 and M82 (Barbieri \etal 1974; Yun, Ho, \& Lo 1995).  From
ground-based images, it is known that its central region contains a single 
bright knot (Price \& Gullixson 1989), which is the dominant feature in the 
FOC image.  The central UV morphology strongly resembles that of NGC~1705 and 
NGC~5253 (Meurer \etal 1995).  Assuming a distance of 3.6 Mpc to M81 (Freedman 
\etal 1994), the central cluster has $L_{2200}$ = 1.0\e{36} \lum\perang, or 
$M_{2200}$ = --10.9 mag (Meurer \etal 1995).  According to Meurer et al.'s 
definition, this cluster would not qualify as an SSC; but like other SSCs, its 
emission is very compact --- the half-light radius measures \lax 0.4 pc. A 
number of fainter clusters are also visible in the frame, the faintest of 
which may be individual O and B stars.


The morphology of NGC~7462, on the other hand is much more complicated.  Many 
compact clusters dot the extent of the large swath of diffuse UV emission 
running parallel to the major axis of the disk.  Assuming a distance of 
13 Mpc (Tully 1988), the clusters have typical sizes of $R_{\rm h}\,\approx$ 
1.5--6 pc and luminosities of $L_{2200}\,\approx$ 1\e{36} \lum\perang.

Although the full analysis of the data base is yet to be completed (Ho \etal 
1996), clusters have been identified in approximately 40\% of the sample.  
This fraction is merely a lower limit, since presumably some of the images 
with no UV emission suffer from high extinction.  The vast 
majority of the galaxies in the survey, strictly speaking, are not considered 
starbursts.  The central regions of a substantial fraction of nearby 
galaxies undergo some level of current star formation (Ho 1995; Ho, Filippenko, 
\& Sargent 1996b), albeit with quite modest star-formation rates.  Moreover, 
the sample objects generally show no outstanding morphological peculiarities 
indicative of recent interactions.  An immediate conclusion that can be 
drawn is that cluster formation is very commonplace and does not 
require very extreme physical conditions.


\section{Star Clusters in Circumnuclear Rings}

S\'ersic \& Pastoriza (1965, 1967) noticed thirty years ago that the 
circumnuclear regions of some barred galaxies have ``hot-spots'' of very 
intense star formation.  It is now accepted that such hot-spots often
delineate a ring-like structure surrounding the nucleus, generally coinciding 
with the location of the inner Lindblad resonance associated with some barred 
spirals.  Gas torqued by the stellar bar loses angular momentum and flows 
inward, and if an inner Lindblad resonance is present, it accumulates in a 
tightly-wound, two-arm spiral or ring-like configuration at the position of 
the resonance (Elmegreen and Friedli, these proceedings).  Large concentrations 
of molecular gas have been detected (e.g., Kenney \etal 1992; Kenney, these 
proceedings), and signs of 
intense star formation are often seen (e.g., Hummel, van der Hulst, \& Keel 
1987; M\'arquez \& Moles 1993).  Hence, it is fair to characterize some 
circumnuclear rings as ``starburst-like,'' even if the globally-averaged rate 
of star formation of the entire galaxy may not be that outstanding.  

{\it HST} images show that young clusters are found in great abundance in 
some circumnuclear rings (Fig. 2).  
In the case of NGC~1097, nearly 90 clusters were identified by Barth \etal 
(1995) in a ring-like structure of diameter $\sim$1 kpc.  Among those for 
which it was possible to perform reliable aperture photometry, the mean 
$R_{\rm h}\,\approx$ 2.5 pc, in accord with the typical half-light radii seen 
in other SSCs.  The clusters have a range of luminosities, the brightest 
having $M_V\,\approx$ --14 mag, with a number between $M_V$ = --13 and --14 
mag.  Numerous clusters are also seen 
in the nuclear ring in NGC~6951, whose large-scale morphology appears very 
similar to the ring in NGC~1097.  The greater distance of this galaxy, 
however, renders photometry of its clusters more difficult.  Barth 
\etal (1995) find $R_{\rm h}\,\leq$ 4 pc for several clusters and $M_V$ 
possibly as luminous as --15 mag, depending on the extinction (which is 
difficult or impossible to measure for individual clusters).  Two other 
examples of nuclear 
rings from Barth \etal (1996) are also shown in Figure 2.  The ring in 
NGC~1019 contains just a single bright cluster, whereas that of NGC~7469 
has a large number.   Note that the apparently strong correlation 
between circumnuclear rings and active galactic nuclei (all four examples 
have either Seyfert or LINER nuclei) is a result of a selection effect; 
the original samples from which these objects were selected largely targeted 
active nuclei.

The FOC UV imaging survey uncovered 5 additional circumnuclear rings 
richly populated with young clusters (Fig. 3).
Although the rings in all but NGC~1079 were previously known from ground-based 
studies, what was not known was that the sites of star formation in the 
rings break up into discrete, compact units (clusters).  In fact, Maoz 
\etal (1996a) find that the clusters constitute a significant fraction 
(15\%--50\%) of the total UV light in these objects.  This estimate is a 
lower limit because there could be fainter clusters undetected at the 
current sensitivities of the snapshot images.  The high detection rate of 
discrete sources implies 
that cluster formation is a significant, if not dominant, mode of star 
formation in these systems.  Meurer \etal (1995) arrived at a similar 
conclusion from their study of 9 starburst systems.  The luminosities of the 
clusters in the rings tend to be lower than those in the sample of 
Meurer et al.  Strictly speaking, they do not qualify as SSCs, but it is 
clear that they are physically similar objects.  The power-law luminosity 
function extrapolated from higher luminosities adequately fits the faint end 
of the distribution (Maoz \etal 1996a).

\section{Dynamical Masses of Super Star Clusters}

One of the central, unanswered questions is whether SSCs are genuine young 
globular clusters.  Thus far, the arguments advanced in favor of this idea 
have been largely inconclusive because the observational evidence is
subject to different interpretations.  Perhaps the most pertinent piece of 
missing information is the mass of the clusters.  The total stellar mass
is very difficult to estimate reliably, since virtually all of the observables 
trace the young, massive stars, which comprise only a small fraction of the 
total mass for a normal IMF.  It would be highly advantageous to bypass this 
complication by obtaining a {\it direct}, model-independent 
measurement of the dynamical mass.  

\begin{figure}
\figurenum{4}
\hbox{
\hskip 0.25truein
\vbox{\hsize 3.0 truein
\psfig{file=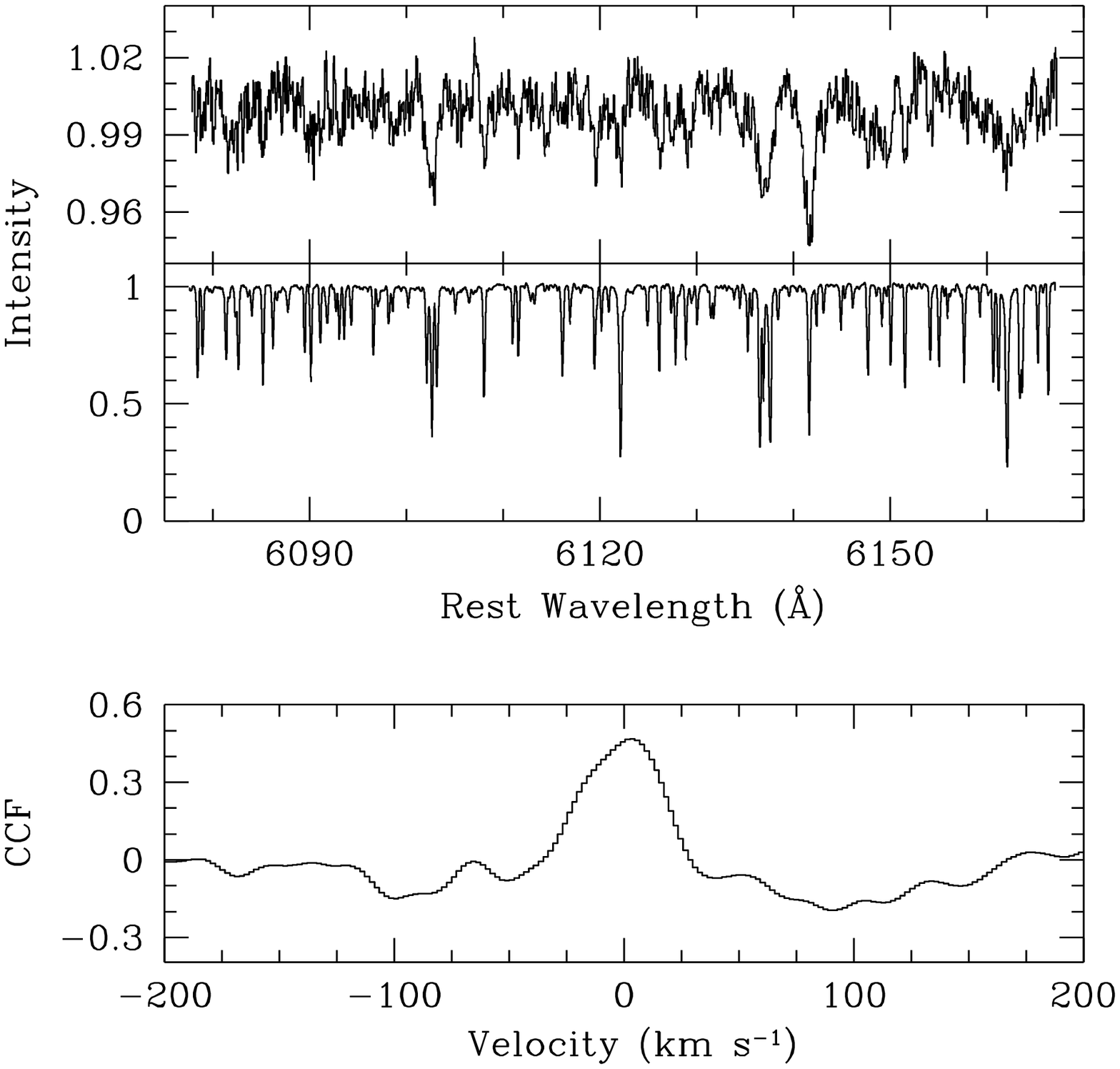,height=3.0truein,width=2.8truein,angle=0}
}
\hskip 0.2truein
\vbox{\hsize 3.0 truein
\psfig{file=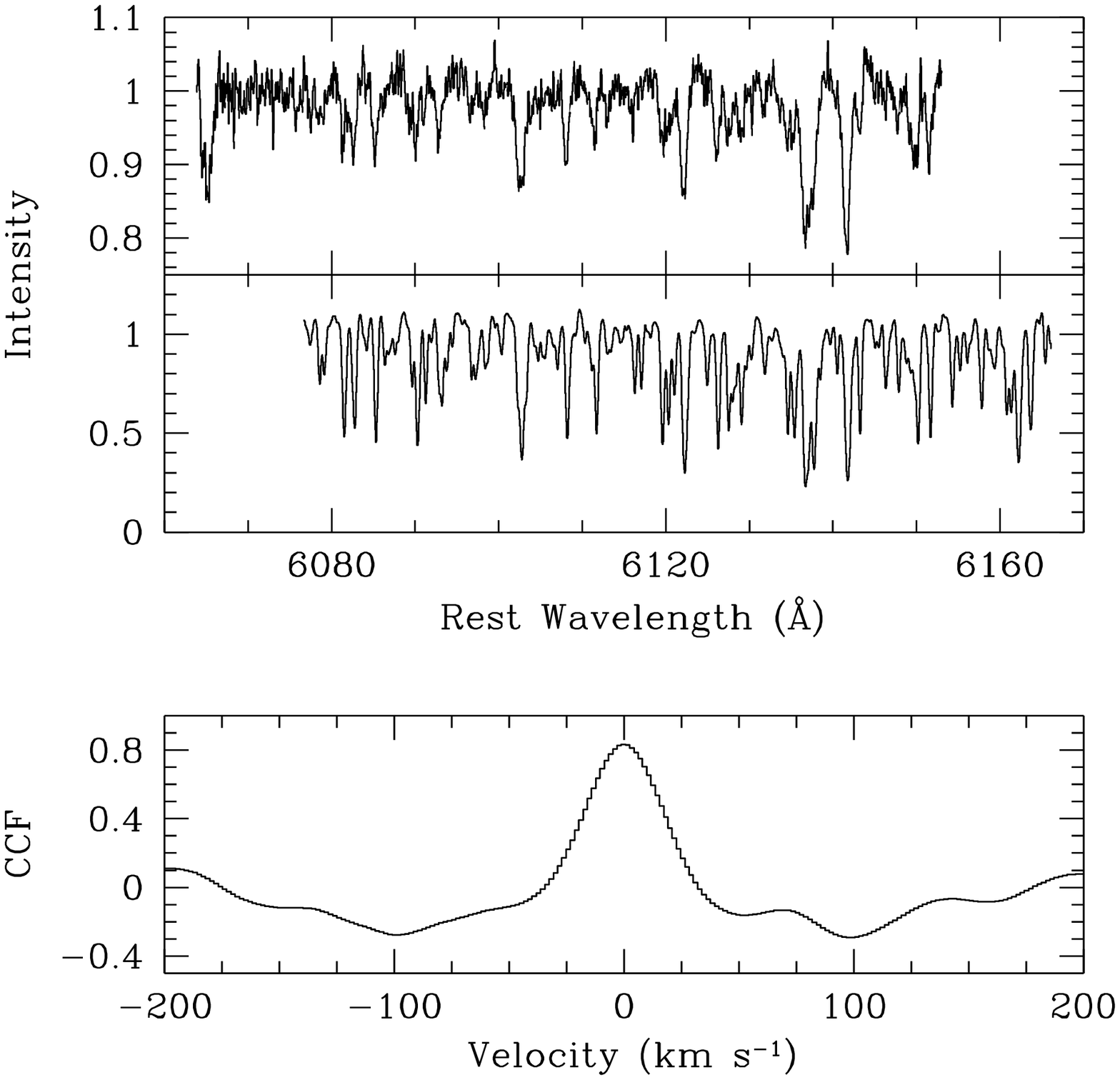,height=3.0truein,width=2.8truein,angle=0}
}
}
\caption{Measurement of the stellar velocity dispersions in cluster ``A'' 
of NGC~1569 ({\it left}; Ho \& Filippenko 1996a) and in the central cluster of 
NGC~1705 ({\it right}; Ho \& Filippenko 1996b).  The top panel in each case 
shows the cluster spectrum, the middle panel the template star used to derive 
the velocity dispersion, and the bottom panel the cross-correlation function 
between the cluster and star.  The width of the main velocity peak of the 
cross-correlation function is related to the velocity dispersion.}\label{fig4}
\end{figure}

Using the HIRES spectrograph mounted on the Keck 10~m telescope, Ho \& 
Filippenko (1996a, b) recently acquired high-dispersion (FWHM = 7.9 \kms) 
optical spectra of the central cluster in NGC~1705 (NGC~1705-1) and of cluster 
``A'' in NGC~1569 (NGC~1569-A), both of which have fairly reliable sizes 
determined from {\it HST} images (O'Connell \etal 1994; Meurer \etal 1995).  
Since these clusters have ages of 10--20 Myr, a 
substantial population of cool supergiants contributes to the integrated 
spectrum at visual wavelengths, and Ho \& Filippenko demonstrate that 
it is feasible to derive the line-of-sight velocity dispersion from the 
weak metal lines detected in both objects.  Figure~\ref{fig4}
illustrates the cross-correlation technique of Tonry \& Davis (1979) applied 
to the two clusters.  

The line-of-sight velocity dispersions are found to be $\sigma_*$ = 
15.7$\pm$1.5 and 11.4$\pm$1.5 \kms\ for NGC~1569-A and NGC~1705-1, 
respectively.  Assuming that the clusters are gravitationally bound (as seems 
quite likely considering their compactness), spherically symmetric, and 
isotropic systems, the dynamical mass is given by $M\,=\,3\sigma_*^2 R/G$ 
according to the virial theorem, where $R$ is the effective gravitational 
radius.  As a first-order approximation, let us assume $R$ = $R_{\rm h}$, the 
half-light radius.  Meurer \etal (1995) recently reanalyzed the images of 
O'Connell \etal (1994) and determined $R_{\rm h}$ = 1.9$\pm$0.2 and 
0.9$\pm$0.2 pc for NGC~1569-A and NGC~1705-1, respectively.  The corresponding 
masses [(3.3$\pm$0.5)\e{5} and (8.2$\pm$2.1)\e{4} \solmass] and mass densities 
(1.1\e{4} and 2.7\e{4} \solmass\ pc$^{-3}$) agree remarkably well with the 
typical values of evolved globular clusters in the Galaxy.  Galactic globular 
clusters have an average mass of 1.9\e{5} \solmass\ (Mandushev, Spassova, \& 
Staneva 1991), and their sizes (and hence densities) are similar to those of 
the young clusters.  Both clusters have absolute visual magnitudes near --14, 
and their estimated ages are 10--20 Myr.  In 10--15 Gyr, they will 
fade by 6 to 7 mag in the $V$ band (e.g., Bruzual \& Charlot 1993), reaching
$M_V$ = --7 to --8 mag, in excellent agreement with the peak of the nearly 
universal luminosity function of globular cluster systems 
($\langle M_V \rangle\,\approx$ --7.3 mag; Harris 1991).  Moreover, if mass 
loss during advanced stages of stellar evolution will not significantly reduce 
the cluster masses, the mass-to-light ratios of NGC~1569-A and NGC~1705-1 
will be $M/L_V$ = 2.5--6.3 and 0.7--1.6 ($M/L_V$)$_{\odot}$, respectively, 
again in reasonable agreement with the observed range for Galactic globular
clusters [0.7--2.9 ($M/L_V$)$_{\odot}$; Mandushev \etal 1991].  
This finding implies that, to a first approximation, the stellar IMF
of the clusters is similar to that of typical globular clusters in the 
Milky Way.

The apparent similarity between the stellar content of SSCs and globular 
clusters has important consequences  for the IMF of starbursts.  In the 
well-known case of M82, Rieke \etal (1980, 1993) have argued that the star 
formation is biased toward high-mass stars; specifically, the IMF evidently 
truncates below $\sim$3 \solmass.  This provocative result, however, has 
been hotly debated (e.g., Scalo 1986, 1987; Zinnecker 1996).  Since SSCs 
appear to be 
the basic building blocks or ``cells'' of star formation in starbursts, and, 
as shown above, at least {\it some} SSCs seem to be genuine young globular 
clusters, the implication is that starburst regions may well have an IMF 
similar to that of globular clusters, which are richly populated with low-mass 
stars (e.g., Paresce \etal 1995).  At this meeting, Moffat, Zinnecker, and 
others have presented evidence that some nearby starburst regions evidently 
also have fairly normal IMFs.

\section{Summary}

1. Compact, young star clusters are very common in a wide range of 
starburst environments.  The formation of bound clusters appears to be a 
major, if not ubiquitous, mode of star formation in starbursts systems.

2. A new {\it HST} imaging survey in the UV bandpass confirms the widespread 
occurance of compact clusters, even in relatively ``normal'' galaxies.  
Star-forming circumnuclear rings often contain rich populations of clusters.

3. The clusters are young (ages ranging from a few to a few hundred Myr), 
compact (half-light radii $\leq$ few parsecs), and have a wide range of 
luminosities.  The masses of the clusters, as inferred from their photometric 
properties, generally lie in the range of 10$^4$--10$^6$ \solmass.
The luminosity function can be represented by a power law with a slope of 
approximately --2.  The most luminous members --- the so-called super star 
clusters --- have luminosities up to 1--2 orders of magnitude higher than 
that of the R136 cluster in 30 Doradus.
 
4. Two nearby examples of super star clusters (in NGC~1569 and NGC~1705) have 
dynamical masses, mass densities, and predicted mass-to-light ratios that 
are virtually indistinguishable from those of evolved globular clusters in 
the Galaxy.  This provides compelling evidence that at least some 
super star clusters truly are present-day analogs of young globular clusters.

\acknowledgments

I thank my principal collaborators Aaron Barth, Alex Filippenko, and 
Dani Maoz for permission to discuss our results prior to publication.  
Aaron Barth and Alex Filippenko provided useful comments on a draft of this 
manuscript.
Deidre Hunter kindly made available some digital images used in my talk.
I am grateful to the scientific organizing committee for inviting me to 
the meeting and the local organizing committee for their warm hospitality that 
made my visit to Mexico so enjoyable.  My research is funded by a postdoctoral 
fellowship from the Harvard-Smithsonian Center for Astrophysics.


\vskip 1.0truein
\centerline {FIGURE CAPTIONS}

\medskip
Fig. 1. --- Sample F220W images from the FOC snapshot survey (Maoz \etal 1996b).

\medskip
Fig. 2. --- Four circumnuclear rings containing compact, young clusters
from {\it HST} images taken at visual wavelengths.  NGC~1097 and NGC~6951
are from Barth \etal (1995), and NGC~1019 and NGC~7469 are from Barth \etal
(1996).  The bright point source south of the nucleus is SN~1992bd.

\medskip
Fig. 3. --- Four of the 5 circumnuclear rings from the FOC UV snapshot survey
(Maoz \etal 1996a) containing compact, young clusters.


\begin{references}

\reference 
Arp, H., \& Sandage, A. 1985, AJ, 90, 1163

\reference 
Ashman, K.~M., \& Zepf, S.~E. 1992, \apj, 384, 50

\reference 
Barbieri, C., Bertola, F., \& di Tullio, G. 1974, \aa, 35, 463

\reference 
Barth, A.~J., Ho, L.~C., Filippenko, A.~V., \& Sargent, W.~L.~W. 1995, \aj,
110, 1009

\reference 
Barth, A. J., Ho, L. C., Filippenko, A. V., Gorjian, V., Malkan,
M., \& Sargent, W. L. W. 1996, in IAU Colloq. 157, Barred Galaxies, ed.\ R. 
Buta, B. G. Elmegreen, \& D. A. Crocker (San Francisco: ASP), 94

\reference 
Benedict, G.~F., \etal 1993, \aj, 105, 1369


\reference 
Bower, G.~A., \& Wilson, A.~S. 1995, \apjs, 99, 543

\reference 
Bruzual A., G., \& Charlot, S. 1993, \apj, 405, 538

\reference 
Burstein, D. 1987, in Nearly Normal Galaxies, ed. S.~M. Faber (New York:
Springer), 47



\reference 
Conti, P.~S. 1991, \apj, 377, 115

\reference 
Conti, P.~S., Leitherer, C., \& Vacca, W.~D. 1996, \apj, 461, L87

\reference 
Conti, P.~S., \& Vacca, W.~D. 1994, \apj, 423, L97



\reference 
Dufour, R.~J., Garnett, D.~R., \& Shields, G.~A. 1988, \apj, 332, 752

\reference 
Faber, S.~M. 1993, in The Globular Cluster-Galaxy Connection, ed. G.~H. Smith
\& J.~P. Brodie (San Francisco: ASP), 601


\reference 
Freedman, W.~L., \etal 1994, \apj, 427, 628

\reference 
Gorjian, V. 1996, \apj, submitted



\reference 
Harris, W.~E. 1991, \annrev, 29, 543


\reference 
Ho, L.~C. 1995, Ph.D. thesis, Univ. of California at Berkeley

\reference 
Ho, L.~C., \etal 1996, in preparation


\reference 
Ho, L.~C., \& Filippenko, A.~V. 1996a, \apj, in press

\reference 
Ho, L.~C., \& Filippenko, A.~V. 1996b, \apj, in press

\reference 
Ho, L.~C., Filippenko, A.~V., \& Sargent, W.~L.~W. 1996a, in IAU Colloq. 157,
Barred Galaxies, ed. R. Buta, B.~G. Elmegreen, \& D.~A. Crocker (San
Francisco: ASP), 188
 
\reference 
Ho, L.~C., Filippenko, A.~V., \& Sargent, W.~L.~W. 1996b, in The
Interplay Between Massive Star Formation, the ISM and Galaxy Evolution, ed.
D. Kunth et al. (Paris: Editions Fronti\'eres), in press

\reference 
Holtzman, J.~A., \etal 1992, \aj, 103, 691

\reference 
Hummel, E., van der Hulst, J.~M., \& Keel, W.~C. 1987, \aa, 172, 32


\reference 
Hunter, D.~A., O'Connell, R.~W., \& Gallagher, III, J.~S. 1994, \aj, 108, 84






\reference
Kenney, J.~D.~P., Wilson, C.~D., Scoville, N.~Z., Devereux, N.~A., \& Young,
J.~S. 1992, \apj, 395, L79

\reference 
Larson, R.~B. 1993, in The Globular Cluster-Galaxy Connection, ed. G.~H. Smith
\& J.~P. Brodie (San Francisco: ASP), 673


\reference 
Leitherer, C., Vacca, W.~D., Conti, P.~S., Filippenko, A.~V., Robert, C., \&
Sargent, W.~L.~W. 1996, \apj, in press


\reference 
Lutz, D. 1991, \aa, 245, 31

\reference 
Mandushev, G., Spassova, N., \& Staneva, A. 1991, \aa, 252, 94

\reference 
Maoz, D., Barth, A.~J., Sternberg, A., Filippenko, A.~V.,
Ho, L.~C., Macchetto, F.~D., Rix, H.-W., \& Schneider, D.~P. 1996a, \aj,
in press

\reference 
Maoz, D., Filippenko, A.~V., Ho, L.~C., Macchetto, F.~D., Rix, H.-W., \&
Schneider, D.~P. 1996b, \apjs, in press

\reference 
M\'arquez, I., \& Moles, M. 1993, \aj, 105, 2090


\reference 
Melnick, J., Moles, M., \& Terlevich, R. 1985, \aa, 149, L24

\reference 
Meurer, G.~R. 1995, \nat, 375, 742


\reference 
Meurer, G.~R., Heckman, T.~M., Leitherer, C., Kinney, A., Robert, C., \&
Garnett, D.~R. 1995, \aj, 110, 2665


\reference 
Meynet, G. 1995, \aa, 298, 767


\reference 
O'Connell, R.~W., Gallagher, J.~S., \& Hunter, D.~A. 1994, \apj, 433, 65

\reference 
O'Connell, R.~W., Gallagher, J.~S., Hunter, D.~A., \& Colley, W.~N. 1995, \apj,
446, L1

\reference 
Paresce, F., de Marchi, G., \& Romaniello, M. 1995, \apj, 440, 216


\reference 
Price, J.~S., \& Gullixson, C.~A. 1989, \apj, 337, 658

\reference 
Rieke, G.~H., Lebofsky, M.~J., Thompson, R.~I., Low, F.~J., \& Tokunaga, A.~T.
1980, \apj, 238, 24
 
\reference 
Rieke, G.~H., Loken, K., Rieke, M.~J., \& Tamblyn, P. 1993, \apj, 412, 99
 
\reference 
Scalo, J.~M. 1986, Fundam. Cosm. Phys., 11, 1

\reference 
Scalo, J.~M. 1987, in Starbursts and Galaxy Evolution, ed. T.~X. Thuan, T.
Montmerle, \& J.~T.~T. Van (Guf sur Yvette: Editions Fronti\'eres), 445

\reference 
Schweizer, F. 1987, in Nearly Normal Galaxies, ed. S.~M. Faber (New York: 
Springer), 18

\reference 
Schweizer, F., \& Seitzer, P. 1993, \apj, 417, L29

\reference 
S\'ersic, J.~L., \& Pastoriza, M. 1965, \pasp, 77, 287

\reference 
S\'ersic, J.~L., \& Pastoriza, M. 1967, \pasp, 79, 152

\reference 
Shaya, E.~J., Dowling, D.~M., Currie, D.~G., Faber, S.~M., \& Groth, E.~J.
1994, \aj, 107, 1675

\reference 
Shlosman, I. (ed.) 1994, Mass Transfer Induced Activity in Galaxies, 
(Cambridge: Cambridge Univ. Press)


\reference 
Storchi-Bergmann, T., Wilson, A.~S., \& Baldwin, J.~A. 1996, \apj, 460, 252

\reference 
Tonry, J., \& Davis, M. 1979, \aj, 84, 1511

\reference 
Tully, R.~B. 1988, Nearby Galaxies Catalog (Cambridge: Cambridge Univ. Press)

\reference 
Vacca, W.~D. 1994, in Violent Star Formation, ed. Tenorio-Tagle
(Cambridge Univ. Press), 297

\reference 
Vacca, W.~D. 1996, in The Interplay Between Massive Star Formation, the ISM
and Galaxy Evolution, ed.  D. Kunth et al. (Paris: Editions Fronti\'eres), in
press

\reference 
Vacca, W.~D., \& Conti, P.~S. 1992, \apj, 401, 543

\reference 
van den Bergh, S. 1995, \nat, 374, 215

\reference 
van den Bergh, S., \& Lafontaine, A. 1984, \aj, 89, 1822

\reference 
van den Bergh, S., Morbey, C., \& Pazder, J. 1991, \apj, 375, 594



\reference 
Whitmore, B.~C., \& Schweizer, F. 1995, \aj, 109, 960

\reference 
Whitmore, B.~C., Schweizer, F., Leitherer, C., Borne, K., \& Robert, C. 1993,
\aj, 106, 1354

\reference 
Yun, M.~S., Ho, P.~T.~P., \& Lo, K.-Y. 1995, \nat, 372, 530

\reference 
Zepf, S.~E., Carter, D., Sharples, R.~M., \& Ashman, K.~M. 1995, \apj, 445,
L19

\reference 
Zinnecker, H. 1996, in The Interplay Between Massive Star Formation, the ISM 
and Galaxy Evolution, ed.  D. Kunth et al. (Paris: Editions Fronti\'eres), in 
press

\end{references}
\end{document}